\documentclass[aps,twocolumn,amssymb,superscriptaddres,pra,reprint]{revtex4-2}
\usepackage[final]{changes}
\usepackage[colorlinks=true,citecolor=blue,linkcolor=blue,urlcolor=blue]{hyperref}
\usepackage{amsmath}
\usepackage{amssymb}
\usepackage{bbm}
\usepackage{xcolor}
\usepackage{graphicx}
\usepackage{dcolumn}
\usepackage{bm}
\usepackage{hyperref}
\usepackage{appendix}
\graphicspath{{Latexpicture/}}

\begin{document}

\title{Single photon isolation and nonreciprocal frequency conversion in atom-waveguide systems}

\author{Jun-Cong Zheng}
\affiliation{Ministry of Education Key Laboratory for Nonequilibrium Synthesis and Modulation of Condensed Matter, Shaanxi Province Key Laboratory of Quantum Information and Quantum Optoelectronic Devices, School of Physics, Xi’an Jiaotong University, Xi’an 710049, China}

\author{Xiao-Wei Zheng}
\affiliation{Ministry of Education Key Laboratory for Nonequilibrium Synthesis and Modulation of Condensed Matter, Shaanxi Province Key Laboratory of Quantum Information and Quantum Optoelectronic Devices, School of Physics, Xi’an Jiaotong University, Xi’an 710049, China}

\author{Xin-Lei Hei}
\affiliation{Ministry of Education Key Laboratory for Nonequilibrium Synthesis and Modulation of Condensed Matter, Shaanxi Province Key Laboratory of Quantum Information and Quantum Optoelectronic Devices, School of Physics, Xi’an Jiaotong University, Xi’an 710049, China}

\author{Yi-Fan Qiao}
\affiliation{Ministry of Education Key Laboratory for Nonequilibrium Synthesis and Modulation of Condensed Matter, Shaanxi Province Key Laboratory of Quantum Information and Quantum Optoelectronic Devices, School of Physics, Xi’an Jiaotong University, Xi’an 710049, China}

\author{Xiao-Yu Yao}
\affiliation{Ministry of Education Key Laboratory for Nonequilibrium Synthesis and Modulation of Condensed Matter, Shaanxi Province Key Laboratory of Quantum Information and Quantum Optoelectronic Devices, School of Physics, Xi’an Jiaotong University, Xi’an 710049, China}

\author{Xue-Feng Pan}
\affiliation{Ministry of Education Key Laboratory for Nonequilibrium Synthesis and Modulation of Condensed Matter, Shaanxi Province Key Laboratory of Quantum Information and Quantum Optoelectronic Devices, School of Physics, Xi’an Jiaotong University, Xi’an 710049, China}

\author{Yu-Meng Ren}
\affiliation{Ministry of Education Key Laboratory for Nonequilibrium Synthesis and Modulation of Condensed Matter, Shaanxi Province Key Laboratory of Quantum Information and Quantum Optoelectronic Devices, School of Physics, Xi’an Jiaotong University, Xi’an 710049, China}

\author{Xiao-Wen Huo}
\affiliation{Ministry of Education Key Laboratory for Nonequilibrium Synthesis and Modulation of Condensed Matter, Shaanxi Province Key Laboratory of Quantum Information and Quantum Optoelectronic Devices, School of Physics, Xi’an Jiaotong University, Xi’an 710049, China}

\author{Peng-Bo Li}
\affiliation{Ministry of Education Key Laboratory for Nonequilibrium Synthesis and Modulation of Condensed Matter, Shaanxi Province Key Laboratory of Quantum Information and Quantum Optoelectronic Devices, School of Physics, Xi’an Jiaotong University, Xi’an 710049, China}

\begin{abstract}
In this work, we utilize a two-level atom and a $\Lambda$-type atom to link two identical waveguides, subsequently extending the model to a giant-atom configuration. Our analytical solutions and numerical simulations demonstrate that this setup can achieve single-photon isolation and nonreciprocal frequency conversion by tuning the atom-waveguide coupling strengths $g_i$, respectively.
We also examine single-photon scattering in the giant-atom model within both the Markovian and non-Markovian regimes. The results reveal that ultranarrow scattering windows are induced by the phases $\phi_1$ and $\phi_2$
under specific conditions, making them well-suited for precise frequency conversion and sensing. Additionally, in the non-Markovian regime, the spectra exhibit irregular polygonal shapes, offering enhanced opportunities for exploring  nonreciprocal frequency conversion in the off-resonant regime. Our work provides a new perspective on achieving optical nonreciprocity at the single-photon level in atom-waveguide systems.

\end{abstract}
\maketitle
\section{introduction}\label{I}

Nonreciprocal photon transmission is a central research topic in chiral quantum optics \cite{lodahl2017chiral}. Platforms like optical diodes \cite{wang2013optical,xia2013ultrabroadband,tokura2018nonreciprocal,tang2019chip,yao2022nonreciprocal,zheng2023few,shen2023tunable} and circulators \cite{riedinger2016non,mahoney2017chip,xia2014reversible,xia2018cavity,lecocq2021control} are widely used in integrated optical circuits for signal routing and processing \cite{kimble2008quantum,jalas2013and}. Beyond one-way photon transmission, nonreciprocal effects also include phenomena such as nonreciprocal photon blockade \cite{PhysRevLett.121.153601,PhysRevA.106.053707,li2019nonreciprocal}, unidirectional quantum amplifiers \cite{PhysRevX.5.021025,PhysRevApplied.7.034031,PhysRevLett.120.023601}, and nonreciprocal bundle emission \cite{PhysRevLett.133.043601}. With advancements in waveguide fabrication and the integration of quantum emitters \cite{PhysRevX.2.011014,RevModPhys.87.347,PhysRevLett.104.203603,goban2014atom}, atom-waveguide systems have gained popularity as a platform for controlling light-matter interactions at ultra-low power levels, down to the single-photon regime. Furthermore, to meet the demands of single-photon frequency conversion \cite{fejer1994nonlinear,wallquist2009hybrid}, various schemes based on atom-waveguide systems have been proposed. These include early explorations of a $\Lambda$-type atom-waveguide system for frequency conversion \cite{PhysRevLett.108.103902,PhysRevA.104.023712} as well as the use of chiral $\Lambda$-type atom-waveguide systems for nonreciprocal frequency conversion \cite{PhysRevResearch.3.043226}. However, achieving nonreciprocal single-photon frequency conversion in nonchiral waveguides remains a significant challenge.

Giant atoms, which challenge the dipole approximation, have been the subject of recent experimental advances \cite{gustafsson2014propagating,satzinger2018quantum,PhysRevLett.124.240402,PhysRevA.103.023710}. These atoms interact with waveguide modes at multiple points separated by large distances. The concept of giant atoms was first theoretically studied in 2014, where the Markovian approximation was used to derive the standard master equation and explore the frequency-dependent behavior \cite{PhysRevA.90.013837}.
Thanks to the phase differences resulting from parametric couplings and direction-dependent phase delays, giant atom systems exhibit distinctive phenomena such as frequency-dependent relaxation rates and Lamb shifts \cite{PhysRevA.90.013837,PhysRevA.104.033710}; unconventional bound states \cite{PhysRevResearch.2.043014,PhysRevLett.126.043602,PhysRevA.101.053855}; and decoherence-free interatomic interactions \cite{PhysRevLett.120.140404,PhysRevResearch.2.043184,PhysRevResearch.2.043070,PhysRevA.105.023712}.
Additionally, non-Markovian retardation effects have become a key area of interest in giant atom systems \cite{andersson2019non,PhysRevResearch.2.043014,PhysRevA.95.053821,PhysRevA.103.053701,PhysRevA.104.033710,PhysRevResearch.4.023198}, where the photon propagation time between different coupling points exceeds the lifetime of the atom.

In this work, we design a nonchiral atom-waveguide system to achieve single-photon isolation and nonreciprocal frequency conversion, respevtively. The setup consists of a two-level atom and a $\Lambda$-type atom simultaneously coupled to two identical waveguides at four coupling points. The transmission of single photons is influenced by the coupling strength. Specifically, when the coupling between the $\vert s \rangle \leftrightarrow \vert e \rangle$ transition of the $\Lambda$-type atom and waveguide $N$ is suppressed, this atom acts as a mirror, reflecting photons incident from waveguide $M$ back into the incident ports. Conversely, photons incident from waveguide $N$ are smoothly transmitted to waveguide $M$ through the two-level atom, as there is no interaction with the three-level atom.
When the aforementioned coupling is not suppressed, nonreciprocal frequency conversion can occur under certain conditions. This frequency conversion arises from the transition between the two lower energy levels of the three-level atom (i.e., from the initial state $\vert g \rangle$ to the final state $\vert s \rangle$) \cite{PhysRevA.109.063709}. The proportion of photons with different wave vectors in waveguide $N$ is fully controlled by the respective coupling strengths, similar to the mechanisms used in generating arbitrary $W$ states \cite{PhysRevLett.127.043604,PhysRevA.105.062408}.
Additionally, we extend the model to a giant atom system and investigate single-photon transmission in both Markovian and non-Markovian regimes. Our research demonstrates that the size of the giant atoms significantly affects the scattering properties of photons. In both regimes, ultranarrow scattering windows are observed near resonance, making them suitable for precise frequency conversion and sensing applications \cite{PhysRevLett.91.243002}. \begin{figure}[htbp]
	\centering
	\resizebox{1\columnwidth}{!}{
		\includegraphics{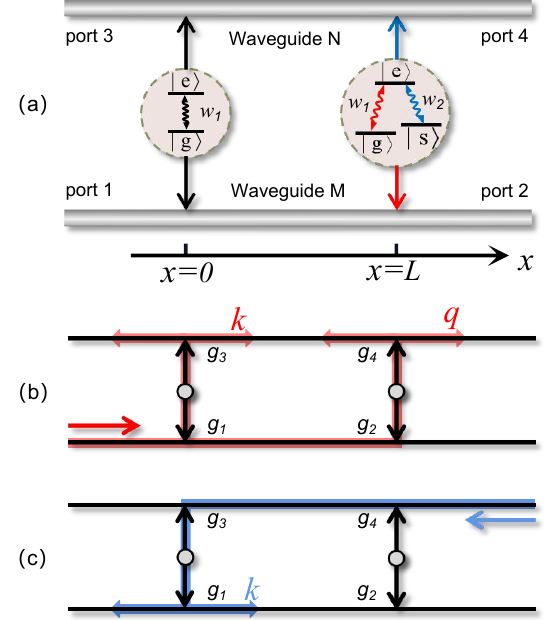}
	}
	\renewcommand\figurename{\textbf{FIG.}}
	\caption[1]{(a) Schematic diagram of the configuration: The two-level atom is coupled to waveguides $M$ and $N$ at $x = 0$ with coupling strengths $g_1$ and $g_3$, respectively. At $x = L$, the $\Lambda$-type atom is coupled to the waveguides, where the transition $\vert g \rangle \leftrightarrow \vert e \rangle$ is coupled to waveguide $M$ (red line) with coupling strength $g_2$, and the transition $\vert s \rangle \leftrightarrow \vert e \rangle$ is coupled to waveguide $N$ (blue line) with coupling strength $g_4$. The transition frequencies between the energy states are $\omega_1$ and $\omega_2$. (b) and (c) show simplified diagrams of the single-photon transport paths with the photon incident from port 1 or port 4.}
	\label{fig.1}
\end{figure}
In the non-Markovian regime, the spectra exhibit polygonal shapes. Notably, for the case where the phase $\phi = \pi$, the transmission probability remains above zero even at non-resonant frequencies.

This work is organized as follows: In Section \ref{II}, we introduce the model for the small atom system, including the corresponding Hamiltonian and the scattering process. Section \ref{IIA} analyzes single-photon isolation in the small atom system, while Section \ref{IIB} focuses on nonreciprocal frequency conversion.
In Section \ref{III}, we extend the model by replacing the small atoms with giant atoms, and investigate the single-photon scattering process in both the Markovian regime (Section \ref{IIIA}) and the non-Markovian regime (Section \ref{IIIB}).
Section \ref{IV} discusses the experimental feasibility of our model, and Section \ref{V} concludes with a summary of the findings.

\section{Small atom system}\label{II}

As shown in Fig.\ref{fig.1}(a), the configuration comprises two infinite linear waveguides, referred to as the lower bus waveguide $M$ and the upper drop waveguide $N$. The waveguides are connected by a two-level atom at $x = 0$ and a $\Lambda$-type  atom at $x = L$. These two atoms have $\omega_1$ and $\omega_s$ as the energies of the states $\vert e \rangle$ and $\vert s \rangle$ with respect to the state $\vert g \rangle$, respectively. Initially, both atoms are prepared in the ground state $\vert g \rangle$.
When a single photon with wave vector $k$ is incident from port 1, as shown in Fig.\ref{fig.1}(b), it will be absorbed by both atoms and then scattered into the waveguides. Considering photon propagation in waveguide $N$, if the photon is transferred by the two-level atom, it undergoes elastic scattering, and the wave vector of the single photon remains unchanged. Conversely, if the photon is transferred by the three-level atom through the transition $\vert s \rangle \leftrightarrow \vert e \rangle$ , it undergoes inelastic scattering with frequency conversion $k \rightarrow q$, where $q = k - \omega_s / v_g$ \cite{Witthaut_2010,PhysRevLett.108.103902,PhysRevA.95.063809}. Notably, the ratio of the wave vector $k$ to $q$ in waveguide $N$ can be adjusted by varying the coupling strengths $g_i$ ($i = 1, 2, 3, 4$).
Similarly, when the same photon is incident from port 4, as shown in Fig.\ref{fig.1}(c), the transition $\vert s \rangle \leftrightarrow \vert e \rangle$ is blocked. Consequently, the photon will only be absorbed by the two-level atom, and thus the frequency of the photon propagation in waveguide $M$ remains unchanged. In summary, this setup allows for the selection of appropriate $g_i$ values to achieve nonreciprocal frequency conversion between waveguide $M$ and waveguide $N$.

The system’s Hamiltonian can be represented as (assuming $\hbar=1$).

\begin{equation}
	\begin{aligned}
		&H=H_a+H_w+H_{int}, \\
		&H_a=\omega_{1}\sigma_{g1}^+\sigma_{g1}^- +\omega_{1}\sigma_{g2}^+\sigma_{g2}^{-}+\omega_{s}\sigma_{s2}^-\sigma_{s2}^+, \\
		&H_w=\sum_{\alpha=M,N}\int_{-\infty}^{\infty} dx[ L^{\dagger}_{\alpha}(x)(\omega_{\alpha}+iv_g\dfrac{\partial}{\partial x})L_{\alpha}(x)\\
		&~~~~~~~~~+R^{\dagger}_{\alpha}(x)(\omega_{\alpha}-iv_g\dfrac{\partial}{\partial x})R_{\alpha}(x)], \\		
		&H_{\textrm{int}}=\sum_{\beta=L,R}\int_{-\infty}^{\infty} dx\lbrace[\delta(x) g_1 \sigma_{g1}^- +\delta(x-L)g_2 \sigma_{g2}^-]\beta_{M}^{\dagger}(x)\\
		&~~~~~~~~~[\delta(x) g_3 \sigma_{g1}^- +\delta(x-L)g_4 \sigma_{s2}^-]\beta_{N}^{\dagger}(x)+H.c.\rbrace.
	\end{aligned}
\end{equation}\label{eq1} 
Here, $H_a$ represents the free Hamiltonian of the atoms, where $\sigma_{\gamma}^{+}$ and $\sigma_{\gamma}^{-}$ ($\gamma = g1, g2, s2$) denote the raising and lowering operators of the atom, respectively. $H_w$ is the Hamiltonian of the waveguides, with $L^{\dagger}_{\alpha}(x)$ [$R^{\dagger}_{\alpha}(x)$] being the creation operator for the right-moving (left-moving) waveguide mode at position $x$, and $v_g$ representing the group velocity. $\omega_{\alpha}$ is the chosen frequency, offset from the cutoff frequency, around which the waveguide's dispersion relation can be linearized as $E = \omega_{\alpha} + kv_g$ or $E = \omega_{N} + qv_g + \omega_s$ \cite{PhysRevLett.95.213001,PhysRevA.79.023837}.
$H_{\textrm{int}}$ describes the interactions between the atomic transitions and the waveguide modes. In this work, we assume $\omega_{M} = \omega_{N} = \omega_{0}$. As discussed in Ref. \cite{PhysRevA.94.063817}, the two photonic reservoirs in our system could represent two distinct, uncoupled modes propagating within the same waveguide. We further assume that the two lower states $\vert g \rangle$ and $\vert s \rangle$ of the three-level atom are quasidegenerate, such that both transition frequencies $\omega_1$ and $\omega_2$ lie within the linearized region around $\omega_0$. The case of nondegenerate lower states, where the group velocities corresponding to the two transition frequencies differ, has been explored in Ref. \cite{PhysRevResearch.3.043226}. The results indicated that this difference has no significant impact on the overall outcomes.

We delve into the single-photon scattering process, which can be elucidated by expressing the single-excitation eigenstate as

\begin{eqnarray}\label{eq2} 
\vert \psi \rangle &=&\sum_{\beta=L,R}\int_{-\infty}^{\infty} dx \lbrace c_{\beta,Ns}(x)\beta_{N}^{\dagger}(x)\vert 0,gs\rangle\nonumber \\&&+[c_{\beta ,M}(x)\beta_M^{\dagger}(x)+c_{\beta,Ng}(x)\beta_{N}^{\dagger}(x)]\vert 0,gg\rangle\rbrace\nonumber \\&&+u_{e1}\vert 0,eg\rangle+u_{e2}\vert 0,ge\rangle,
\end{eqnarray}
$c_{\beta,M}(x)$, $c_{\beta,Ng}(x)$, and $c_{\beta,Ns}(x)$ represent the probability amplitudes of finding a photon in the waveguides at position $x$ with the atoms ultimately in states $\vert 0, gg \rangle$ and $\vert 0, gs \rangle$, respectively. $u_{e1}$ and $u_{e2}$ are the probability amplitudes of the atoms being in the excited states $\vert 0, eg \rangle$ and $\vert 0, ge \rangle$, respectively. By deriving the stationary Schrödinger equations for the amplitudes from the eigenvalue equation $H\vert \psi \rangle = E\vert \psi \rangle$, one can obtain

\begin{equation}\label{eq3}
	\begin{aligned}
	&E c_{L,M}(x)=(\omega_0+iv_g\dfrac{\partial}{\partial x})c_{L,M}(x)+g_1 \delta(x)u_{e1}\\
	&~~~~~~~~~~~~~~~~+g_2 \delta(x)u_{e2},\\
	&E c_{R,M}(x)=(\omega_0-iv_g\dfrac{\partial}{\partial x})c_{R,M}(x) +g_1 \delta(x)u_{e1}\\
	&~~~~~~~~~~~~~~~~+g_2 \delta(x)u_{e2},\\
	&E c_{L,Ng}(x)=(\omega_0+iv_g\dfrac{\partial}{\partial x})c_{L,Ng}(x) +g_3 \delta(x)u_{e1},\\
	&E c_{R,Ng}(x)=(\omega_0-iv_g\dfrac{\partial}{\partial x})c_{R,Ng}(x) +g_3 \delta(x)u_{e1},\\
	&E c_{L,Ns}(x)=(\omega_0+\omega_{s}+iv_g\dfrac{\partial}{\partial x})c_{L,Ns}(x) +g_4 \delta(x)u_{e2},\\
	&E c_{R,Ns}(x)=(\omega_0+\omega_{s}-iv_g\dfrac{\partial}{\partial x})c_{R,Ns}(x) +g_4 \delta(x)u_{e2},\\	
	&\Delta u_{e1}=g_1 [c_{L,M}(0)+c_{R,M}(0)]+g_3 
	[c_{L,Ng}(0)+c_{R,Ng}(0)],\\
	&\Delta u_{e2}=g_2 [c_{L,M}(0)+c_{R,M}(0)]+g_4 [c_{L,Ns}(0)+c_{R,Ns}(0)],
	\end{aligned}
\end{equation}
where $\Delta=E-\omega_{1}$ is the detuning between the propagating
photons and the transition $\vert g \rangle \leftrightarrow \vert e \rangle$.

For the first case, where the photon is incident from port 1, the wave functions are given as follows:

\begin{equation}\label{eq4}
\begin{aligned}
&c_{R,M}(x)=\lbrace\Theta(-x)+A_0[\Theta(x)-\Theta(x-L)]\\
&~~~~~~~~~~~~~~~+t_{2}\Theta(x-L)\rbrace e^{ikx},\\
&c_{L,M}(x)=\lbrace r_{1}\Theta(-x)+B_0[\Theta(x)-\Theta(x-L)]\rbrace e^{-ikx},\\
&c_{R,Ng}(x)=t_{4,g}\Theta(x)e^{ikx},\\
&c_{L,Ng}(x)=t_{3,g}\Theta(-x)e^{-ikx},\\
&c_{R,Ns}(x)=t_{4,s}\Theta(x-L)e^{iqx},\\
&c_{L,Ns}(x)=t_{3,s}\Theta(L-x)e^{-iqx},\\
\end{aligned}
\end{equation}
where $\Theta(x)$ is the Heaviside step function; $r_1$ and $t_2$ are the reflection and transmission coefficients on waveguide $M$. $t_{3,g}$ and $t_{4,g}$ ($t_{3,s}$ and $t_{4,s}$) are the transfer coefficients for the elastic (inelastic) scattering processes to port 3 and port 4, respectively. Here, $A_0$ and $B_0$ represent the probability amplitudes of finding right-moving and left-moving photons with wave vector $k$ in the region $0 < x < L$, respectively. By substituting Eq. (\ref{eq4}) into Eq. (\ref{eq3}), we obtain the transmission and reflection amplitudes.

\begin{equation}\label{eq5}
\begin{aligned}
&t_{3,g}=t_{4,g}\\
&=\dfrac{\sqrt{\Gamma_1\Gamma_3}[\Gamma_4-i\Delta-(e^{2i\phi_a}-1)\Gamma_2]}{e^{2i\phi_a}\Gamma_1\Gamma_2-(\Gamma_1+\Gamma_3-i\Delta)(\Gamma_2+\Gamma_4-i\Delta)},\\
&t_{3,s}=t_{4,s}\\
&=\dfrac{e^{i(\phi_a+\phi_b)}\sqrt{\Gamma_2\Gamma_4}(\Gamma_3-i\Delta)}{e^{2i\phi_a}\Gamma_1\Gamma_2-(\Gamma_1+\Gamma_3-i\Delta)(\Gamma_2+\Gamma_4-i\Delta)},\\
&r_{1}=\dfrac{\Gamma_1 (\Gamma_2+\Gamma_4-i\Delta)-e^{2i\phi_a}\Gamma_2(\Gamma_1-\Gamma_3+i\Delta)}{e^{2i\phi_a}\Gamma_1\Gamma_2-(\Gamma_1+\Gamma_3-i\Delta)(\Gamma_2+\Gamma_4-i\Delta)},\\
\end{aligned}
\end{equation}
where $\Gamma_n = g_n^2 / v_g$ ($n = 1, 2, 3, 4$) are the radiative decay rates. $\phi_a = kL$ and $\phi_b = qL$ are the phases accumulated by photons between the coupling points of the two atoms. We verify that $\vert r_{1} \vert^2 + \vert t_{2} \vert^2 + \vert t_{3,g} \vert^2 + \vert t_{4,g} \vert^2+ \vert t_{3,s} \vert^2 + \vert t_{4,s} \vert^2 = 1$ (the analytic expression for $t_{2}$ is not shown here), which ensures the conservation of probability for the incident photon. Specifically, when $L = 0$, the two atoms overlap, and Eq. (\ref{eq5}) simplifies to

\begin{equation}\label{eq6}
\begin{aligned}
&t_{3,g}=t_{4,g}=\dfrac{-\sqrt{\Gamma_1\Gamma_3}(\Gamma_4-i\Delta)}{\Gamma_2(\Gamma_3-i\Delta)+(\Gamma_1+\Gamma_3-i\Delta)(\Gamma_4-i\Delta)},\\
&t_{3,s}=t_{4,s}=\dfrac{-\sqrt{\Gamma_2\Gamma_4}(\Gamma_3-i\Delta)}{\Gamma_2(\Gamma_3-i\Delta)+(\Gamma_1+\Gamma_3-i\Delta)(\Gamma_4-i\Delta)},\\
&r_{1}=\dfrac{\Gamma_2\Gamma_3+\Gamma_1\Gamma_4-i(\Gamma_1+\Gamma_2)\Delta}{\Gamma_2(\Gamma_3-i\Delta)+(\Gamma_1+\Gamma_3-i\Delta)(\Gamma_4-i\Delta)}.
\end{aligned}
\end{equation}

Similarly, for the second case, where the photon is incident from port 4 and $L = 0$, the wave functions are given as follows:

\begin{figure}[htbp]
	\centering
	\resizebox{0.82\columnwidth}{!}{
		\includegraphics{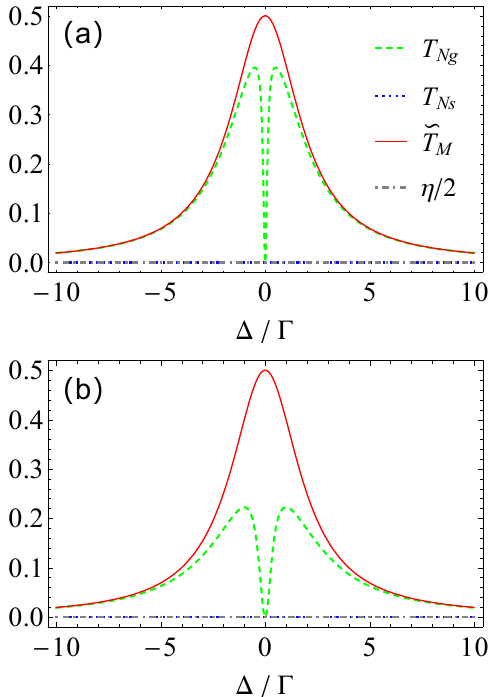}
	}
	\renewcommand\figurename{\textbf{FIG.}}
	\caption[2]{The spectra of the single-photon transfer rates $T_{Ng}$ (green dashed line), $T_{Ns}$ (blue dotted line), $\tilde{T}_{M}$ (red solid line), and half of the proportionality factor $\eta/2$ (gray dot-dashed line) are plotted as functions of detuning $\Delta$. In panel (a), $\Gamma_1 = \Gamma_3 = \Gamma$ and $\Gamma_2 = 0.25\Gamma$; in panel (b), $\Gamma_1 = \Gamma_2 = \Gamma_3 = \Gamma$. These spectra are calculated for the parameter $\Gamma_4 = 0$.}
	\label{fig.2}
\end{figure}

\begin{equation}\label{eq7}
\begin{aligned}
&c_{R,M}(x)=\tilde{t}_{2} \Theta(x) e^{ikx},\\
&c_{L,M}(x)=\tilde{t}_{1} \Theta(-x) e^{-ikx},\\
&c_{R,Ng}(x)=\tilde{r}_{4,g}\Theta(x)e^{ikx},\\
&c_{L,Ng}(x)=[\Theta(x)+\tilde{t}_{3,g}\Theta(-x)]e^{-ikx},\\
&c_{R,Ns}(x)=0,\\
&c_{L,Ns}(x)=0,\\
\end{aligned}
\end{equation}
the transfer coefficients on waveguide $M$ in this case can then be determined as

\begin{equation}
 \tilde{t}_{1}=\tilde{t}_{2}=\dfrac{-\sqrt{\Gamma_1\Gamma_3}}{\Gamma_1+\Gamma_3-i\Delta}.
\end{equation}\label{eq8}

In the following discussion, we assume $T_{Ng} = \vert t_{3,g} \vert^2 + \vert t_{4,g} \vert^2$ ($T_{Ns} = \vert t_{3,s} \vert^2 + \vert t_{4,s} \vert^2$) are the elastic (inelastic) transfer rates on waveguide $N$, and $\eta = T_{Ns} / (T_{Ng} + T_{Ns})$ represents the proportion of frequency conversion occurring on waveguide $N$. $\tilde{T}_{M} = \vert \tilde{t}_{1} \vert^2 + \vert \tilde{t}_{2} \vert^2$ denotes the transfer rates on waveguide $M$.

\begin{figure}[htbp]
	\centering
	\resizebox{1.03\columnwidth}{!}{
		\includegraphics{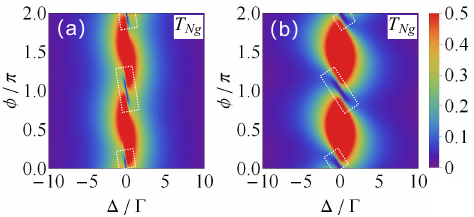}
	}
	\renewcommand\figurename{\textbf{FIG.}}
	\caption[3]{The spectra of the single-photon transfer rate $T_{Ng}$ are plotted as functions of detuning $\Delta$ and phase $\phi_a = \phi$. In panel (a), $\Gamma_1 = \Gamma_3 = \Gamma$ and $\Gamma_2 = 0.25\Gamma$; in panel (b), $\Gamma_1 = \Gamma_2 = \Gamma_3 = \Gamma$. These spectra are calculated for the parameter $\Gamma_4 = 0$.}
	\label{fig.3}
\end{figure}

\subsection{single photon islation}\label{IIA}

From the scattering coefficients calculated in Eq. (\ref{eq6}), it is straightforward to see that when we set $\Gamma_4 = 0$ at resonance (i.e., $\Delta = 0$), the configuration functions as a single-photon isolator. In this scenario, there always exists $T_{Ng} = T_{Ns} = 0$ and $\tilde{T}_{M} \neq 0$. The single photon is only allowed to transfer from waveguide $N$ to waveguide $M$. When the photon is incident from waveguide $M$, there is $r_1 = 1$, indicating that the atom acts as a mirror, blocking photon propagation \cite{Shen:05,PhysRevLett.101.180404,PhysRevLett.100.093603}. In this work, we focus solely on the transfer rates, which are sufficient to capture the essential physics of our models.

Figure \ref{fig.2} illustrates the transfer rates $T_{Ng}$, $T_{Ns}$, $\tilde{T}_{M}$, and half of the proportionality factor $\eta/2$ as functions of detuning $\Delta$ when $L = 0$. For the case of $\Gamma_2 = 0.25\Gamma$ shown in Fig. \ref{fig.2}(a), near resonance, photons can travel from waveguide $N$ to waveguide $M$ without obstruction, while they are blocked in the reverse direction. In this scenario, only photons with wave vector $k$ are present in the system since the atomic transition $\vert s \rangle \leftrightarrow \vert e \rangle$ is decoupled from the waveguide, preventing frequency conversion during the scattering process. In Fig. \ref{fig.2}(b), increasing the decay rate to $\Gamma_2 = \Gamma$ enhances the competition of the three-level atom for the incident photons across the entire bandwidth, resulting in most photons being reflected back to the incident port.

When $L \neq 0$, the phase generated by the relative position between the two atoms also affects the transmission of photons. In Fig. \ref{fig.3}, we plot the spectra of the single-photon transfer rate $T_{Ng}$ as functions of detuning $\Delta$ and phase $\phi_a = \phi$. It is evident that $T_{Ng}$ varies periodically with $\phi_a$. Although the system achieves a higher transfer probability near resonance, there is always a photon-blocking region when $\phi_a$ is an integer multiple of $\pi$ (highlighted by the white rectangle with dotted lines). Additionally, as the decay rate $\Gamma_2$ increases, the photon-blocking region enlarges, which is consistent with the results discussed in Fig. \ref{fig.2}.

\begin{figure}[htbp]
	\centering
	\resizebox{0.8\columnwidth}{!}{
		\includegraphics{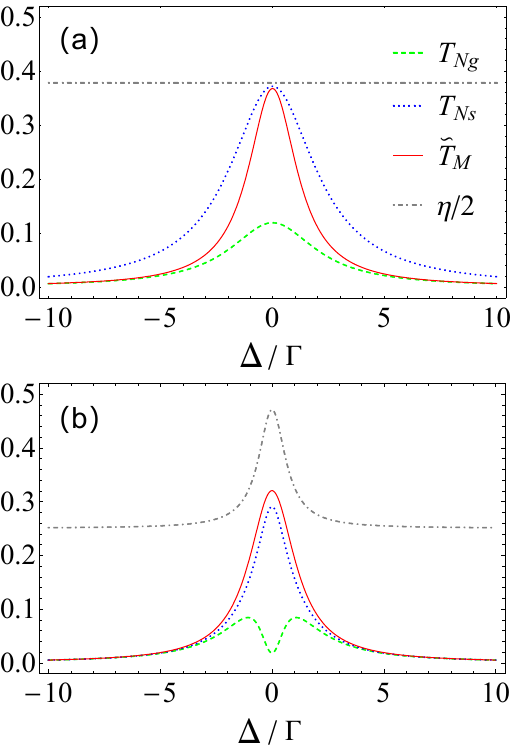}
	}
	\renewcommand\figurename{\textbf{FIG.}}
	\caption[4]{The spectra of the single-photon transfer rates $T_{Ng}$ (green dashed line), $T_{Ns}$ (blue dotted line), $\tilde{T}_{M}$ (red solid line), and half of the proportionality factor $\eta/2$ (gray dot-dashed line) are plotted as functions of detuning $\Delta$. In panel (a), $\Gamma_2 = \Gamma_3 = \Gamma_4 = \Gamma$, and $\Gamma_1 = 0.32 \Gamma$; in panel (b), $\Gamma_2 = \Gamma_3 = \Gamma$, and $\Gamma_1 = \Gamma_4 = 0.25 \Gamma$.}
	\label{fig.4}
\end{figure}

\subsection{nonreciprocal frequency conversion}\label{IIB}

As discussed earlier, the configuration can achieve nonreciprocal frequency conversion by adjusting the coupling strength $g_i$ to increase the probability of inelastic scattering processes. Specifically, increasing the ratio of coupling strength $g_2 / g_1$ enhances the probability that a photon is absorbed by the three-level atom when it is incident from waveguide $M$, then scattered through the atomic transition $\vert s \rangle \leftrightarrow \vert e \rangle$, accompanied by frequency conversion $k\rightarrow q$. On the other hand, to ensure that a photon incident from waveguide $N$ can be scattered through the two-level atom to waveguide $M$ simultaneously, the coupling strength $g_1$ cannot be set too small.

\begin{figure}[htbp]
	\centering
	\resizebox{1\columnwidth}{!}{
		\includegraphics{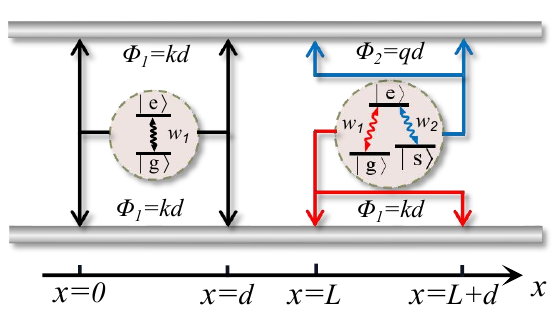}
	}
	\renewcommand\figurename{\textbf{FIG.}}
	\caption[5]{The giant-atom version of the system in Fig. \ref{fig.1} (a) features a two-level atom with atom-waveguide couplings at both points $x = 0$ and $x = d$, while the three-level atom has atom-waveguide couplings at both points $x = L$ and $x = L + d$. Other characteristics of the system remain unchanged.}
	\label{fig.5}
\end{figure}

In Fig. \ref{fig.4}, we present two sets of simulations to investigate nonreciprocal frequency conversion. Specifically, in Fig. \ref{fig.4}(a), the decay rates are set as $\Gamma_2 = \Gamma_3 = \Gamma_4 = \Gamma$ and $\Gamma_1 = 0.32 \Gamma$. The spectrum reveals that, at resonance, the photon transfer probabilities are $T_{Ng} = 0.12$, $T_{Ns} = \tilde{T}_{M} = 0.37$, and the proportionality factor is $\eta \approx 0.76$. At this point, there are nonzero probabilities for photons with wave vectors $k$ and $q$ being transferred within waveguide $N$. Notably, the majority of the photons undergo frequency conversion, shifting from $k$ to $q$ during the scattering process. Additionally, some photons can be reversibly transferred to waveguide $M$ without a frequency change, making this process effectively approximate nonreciprocal frequency conversion.
In Fig. \ref{fig.4}(b), the decay rates are modified to $\Gamma_2 = \Gamma_3 = \Gamma$ and $\Gamma_1 = \Gamma_4 = 0.25 \Gamma$. At resonance, the updated photon transfer probabilities are $T_{Ng} = 0.02$, $T_{Ns} = 0.3$, $\tilde{T}_{M} = 0.32$, and $\eta = 0.94$. Here, the probability of photons with wave vector $k$ remaining in waveguide $N$ after scattering is nearly zero. The process achieves higher purity of frequency conversion, albeit with a slightly lower transfer rate compared to the case shown in Fig. \ref{fig.4}(a). Therefore, selecting appropriate parameters is critical for optimizing the desired functionality in this configuration.

It is evident that, whether for single-photon isolation in section \ref{IIA} or nonreciprocal frequency conversion in section \ref{IIB}, the maximum transmission probability between waveguides does not exceed 0.5. To improve efficiency, consider replacing the infinite waveguide with a semi-infinite waveguide to overcome this limitation. Details can be found in the \hyperref[Appendix]{Appendix}.

\section{Ginat atom system}\label{III}

As shown in Fig. \ref{fig.5}, we replace the small atoms in Fig. \ref{fig.1} (a) with giant atoms, which generate two distinct phases across four regions. In this section, to simplify the calculations, we assume that the positions of the two atoms coincide (i.e., $L = 0$) and that the coupling strengths of the two arms for each atom on the same waveguide are equal. Therefore, the interaction Hamiltonian becomes

\begin{eqnarray}
	H_{\textrm{int}}&=&\sum_{\beta=L,R}\int_{-\infty}^{\infty} dx[\delta(x)+\delta(x-d)][(g_1 \sigma_{g1}^- \nonumber \\&&+g_2 \sigma_{g2}^-)\beta_{M}^{\dagger}(x)+
	(g_3 \sigma_{g1}^- +g_4 \sigma_{s2}^-)\beta_{N}^{\dagger}(x)+H.c.].
\end{eqnarray}\label{eq9}

When a photon is incident from port 1, the wave functions are given by

\begin{equation}\label{eq10}
	\begin{aligned}
		&c_{R,M}(x)=\lbrace\Theta(-x)+A[\Theta(x)-\Theta(x-d)]\\
		&~~~~~~~~~~~~~~+t_{2}\Theta(x-d)\rbrace e^{ikx},\\
		&c_{L,M}(x)=\lbrace r_{1}\Theta(-x)+B[\Theta(x)-\Theta(x-d)]\rbrace e^{-ikx},\\
		&c_{R,Ng}(x)=\lbrace C[\Theta(x)-\Theta(x-d)]+ t_{4,g}\Theta(x-d)\rbrace e^{ikx},\\
		&c_{L,Ng}(x)=\lbrace t_{3,g}\Theta(-x)+D[\Theta(x)-\Theta(x-d)]\rbrace e^{-ikx},\\
		&c_{R,Ns}(x)=\lbrace E[\Theta(x)-\Theta(x-d)]+t_{4,s}\Theta(x-d)\rbrace e^{iqx},\\
		&c_{L,Ns}(x)=\lbrace t_{4,s}\Theta(-x)+F[\Theta(x)-\Theta(x-d)]\rbrace e^{-iqx},\\
	\end{aligned}
\end{equation}
and when the photon is incident from port 4, the wave functions are given by

\begin{equation}\label{eq11}
\begin{aligned}
&c_{R,M}(x)=\lbrace\tilde{t}_{2} \Theta(x-d)+\tilde{A}[\Theta(x)-\Theta(x-d)] \rbrace e^{ikx},\\
&c_{L,M}(x)=\lbrace \tilde{B}[\Theta(x)-\Theta(x-d)]+\tilde{t}_{1} \Theta(-x) \rbrace e^{-ikx},\\
&c_{R,Ng}(x)=\lbrace\tilde{r}_{4,g}\Theta(x-d)+\tilde{C}[\Theta(x)-\Theta(x-d)]\rbrace e^{ikx},\\
&c_{L,Ng}(x)=\lbrace\Theta(x-d)+\tilde{D}[\Theta(x)-\Theta(x-d)]\\
&~~~~~~~~~~~~~~+\tilde{t}_{3,g}\Theta(-x)\rbrace e^{-ikx},\\
&c_{R,Ns}(x)=0,\\
&c_{L,Ns}(x)=0,\\
\end{aligned}
\end{equation}
$A$, $B$, $\tilde{A}$, and $\tilde{B}$, as well as $C$, $D$, $\tilde{C}$, and $\tilde{D}$ ($E$, $F$), are the probability amplitudes for finding photons with wave vector $k$ ($q$) in the region $0 < x < d$ on waveguides $M$ and $N$, respectively. The transmission coefficients can then be determined as

\begin{equation}\label{eq12}
\begin{aligned}
&t_{3,g}=t_{4,g}\\
&=\dfrac{-(1+e^{i\phi_1})^2 \chi_2 \sqrt{\Gamma_1\Gamma_3}}
{2(1+e^{i\phi_1})\chi_1\Gamma_2+\chi_2[2(\Gamma_1+\Gamma_3)(1+e^{i\phi_1})-i\Delta]},\\	
&t_{3,s}=t_{3,s}\\
&=\dfrac{-(1+e^{i\phi_1})(1+e^{i\phi_2}) \chi_1 \sqrt{\Gamma_2\Gamma_4}}
{2(1+e^{i\phi_1})\chi_1\Gamma_2+\chi_2[2(\Gamma_1+\Gamma_3)(1+e^{i\phi_1})-i\Delta]},\\
&\tilde{t}_{1}=\tilde{t}_{2}=\dfrac{-\sqrt{\Gamma_1\Gamma_3}e^{-2i\phi_1}(1+e^{i\phi_1})^2}{2(\Gamma_1+\Gamma_3)(1+e^{i\phi_1})-i\Delta},
\end{aligned}
\end{equation}
with
\begin{equation}\label{eq13}
\begin{aligned}
&\chi_1=2(1+e^{i\phi_1})\Gamma_3-i\Delta,\\	
&\chi_2=2(1+e^{i\phi_2})\Gamma_4-i\Delta,\\
\end{aligned}
\end{equation}
\begin{figure}[htbp]
	\centering
	\resizebox{1\columnwidth}{!}{
		\includegraphics{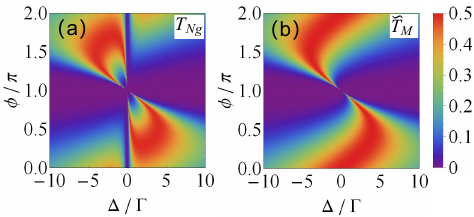}
	}
	\renewcommand\figurename{\textbf{FIG.}}
	\caption[6]{In the Markovian regime, the spectra of the single-photon transfer rate $T_{Ng}$ (panel (a)) and $\tilde{T}_{M}$ (panel (b)) are shown as functions of detuning $\Delta$ and phase $\phi_1^{'} = \phi$. These spectra are calculated with the parameters $\Gamma_1 = \Gamma_3 = \Gamma$, $\Gamma_2 = 0.25 \Gamma$, and $\Gamma_4 = 0$.}
	\label{fig.6}
\end{figure}
where $\phi_1=k d=(\omega_1-\omega_0+\Delta)\tau$, $\phi_2=q d=(\omega_1-\omega_s-\omega_0+\Delta)\tau$, and $\tau=d/v_g$ is the corresponding propagation time between the coupling points. we redefine $\phi_1=\phi_1^{'}+\tau\Delta$ and $\phi_2=\phi_2^{'}+\tau\Delta$ with $\phi_1^{'}=(\omega_1-\omega_0)\tau$ and $\phi_2^{'}=(\omega_1-\omega_s-\omega_0)\tau$, such that both $\phi_1$ and $\phi_2$ divide into a constant part and a $\Delta$-dependent part.  This allows us to study single-photon scattering in both the Markovian and non-Markovian regimes, depending on whether the propagation time $\tau$ is negligible or not.

\subsection{Markovian regime}\label{IIIA}

\begin{figure*}[htbp]
	\centering
	\resizebox{1.8\columnwidth}{!}{
		\includegraphics{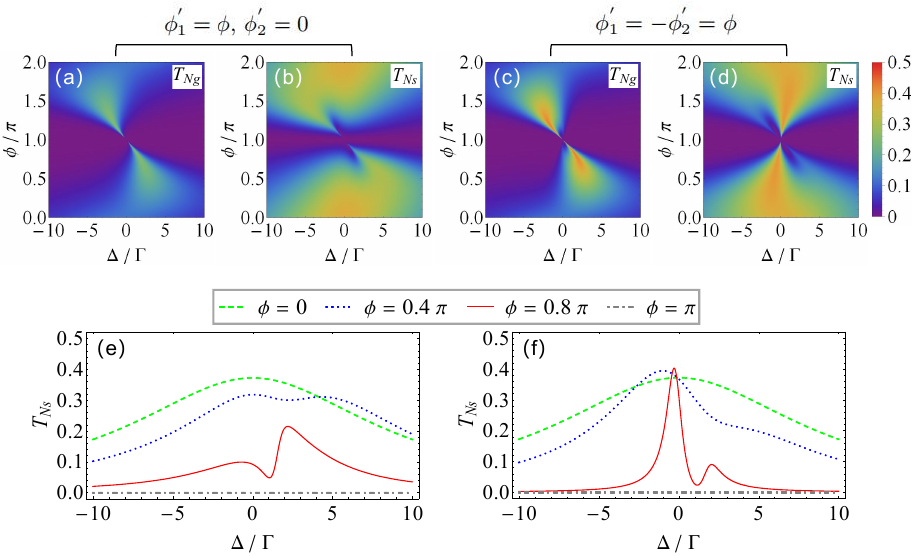}
	}
	\renewcommand\figurename{\textbf{FIG.}}
	\caption[7]{In the Markovian regime, the spectra of the single-photon transfer rate $T_{Ng}$ are shown in panels (a) and (c), and $T_{Ns}$ are shown in panels (b) and (d), as functions of detuning $\Delta$ and phase $\phi$. These spectra are calculated with $\phi_1^{'} = \phi$ and $\phi_2^{'} = 0$ in panels (a) and (b), and with $\phi_1^{'} = -\phi_2^{'} = \phi$ in panels (c) and (d). Panels (e) and (f) depict the profiles of $T_{Ns}$ for different phase values, corresponding to the cases shown in panels (b) and (d), respectively. Other parameters are set to $\Gamma_1 = 0.32 \Gamma$ and $\Gamma_2 = \Gamma_3 = \Gamma_4 = \Gamma$.}
	\label{fig.7}
\end{figure*}

When the parameters satisfy the condition $\tau \sum_i \Gamma_i \ll 1$ \cite{Longhi:20, PhysRevA.95.053821}, the transmission of a single photon can be considered a Markovian process. Specifically, the second part of $\phi_1$ and $\phi_2$ is small enough to be neglected (i.e., $\tau \Delta \approx 0$). By applying the Taylor expansion to the first order of $\tau$, we have $\exp(i \phi_1) \approx \exp(i \phi_1^{'})$ and $\exp(i \phi_2) \approx \exp(i \phi_2^{'})$.

We first consider the case of single-photon isolation  and depict the spectra of the single-photon transfer rate $T_{Ng}$ in Fig.\ref{fig.6}(a) and $\tilde{T}_{M}$ in Fig.\ref{fig.6}(b) as functions of detuning $\Delta$ and phase $\phi_1^{'} = \phi$. Figure \ref{fig.6}(a) shows that the transmission channel for a single photon from waveguide $M$ to waveguide $N$ is always closed near resonance across the entire $2\pi$ phase range. This indicates that the separation $d$ between the two coupling points does not affect the result of the photon being reflected by the atom. As shown in Fig. \ref{fig.6}(b), the large value of the transfer rate $\tilde{T}_{M}$ is concentrated in the $\mathcal{S}$-shaped area. For $\Delta = 0$, when the phase $\phi \in [0, 0.1\pi]$ or $\phi \in [1.9\pi, 2\pi]$, $\tilde{T}_{M} \approx 0.5$, which represents an appropriate parameter range for implementing single-photon nonreciprocal transmission.

We simulate nonreciprocal frequency conversion in the Markovian regime in Fig. \ref{fig.7}. Given that there are two frequencies in the system, we examine the cases of $\phi_1^{'} = \phi$, $\phi_2^{'} = 0$ in Figs. \ref{fig.7}(a) and (b), and $\phi_1^{'} = -\phi_2^{'} = \phi$ in Figs. \ref{fig.7}(c) and (d). In Figs. \ref{fig.7}(a)-(d), the spectra corresponding to $T_{Ng} \neq 0$ and $T_{Ns} \neq 0$ are separated into two regions by the phase $\phi = \pi$. When adjusting the parameter from $\phi_2^{'} = 0$ to $\phi_2^{'} = -\phi$, an effective transmission ultranarrow window is generated for both $T_{Ng}$ and $T_{Ns}$ as $\phi \rightarrow \pi$. Additionally, detailed descriptions of $T_{Ns}$ are shown in Figs. \ref{fig.7}(e) and (f). At resonance, the proportion of frequency conversion is independent of $\phi_1$ and $\phi_2$, remaining constant at $\eta = \Gamma_2 \Gamma_3 / (\Gamma_1 \Gamma_4 + \Gamma_2 \Gamma_3) = 0.76$. For $\phi = 0.8\pi$, when $\phi_2^{'} = 0$, $T_{Ns} = 0.09$, which is not suitable for frequency conversion. However, when $\phi_2^{'} = -\phi$, $T_{Ns} = 0.31$, which is more suitable for frequency conversion. Such ultranarrow scattering windows are expected to have applications in precise frequency conversion and sensing. Furthermore, to achieve nonreciprocal frequency conversion, it is crucial to consider the value of $\tilde{T}_{M}$. It should not be too small to ensure effective transfer of the incident photon from waveguide $N$ to waveguide $M$, as discussed previously.

\subsection{Non-Markovian regime}\label{IIIB}

When $\tau$ is not a small value, the system enters the non-Markovian regime. Similar experimental work includes cases where transmons are coupled with surface acoustic waves \cite{PhysRevA.95.053821, andersson2019non}, and where the separation $d$ between the two coupling points is large enough \cite{kannan2020waveguide}. In such situations, the second parts of $\phi_1$ and $\phi_2$ cannot be neglected, and $\phi_1$ and $\phi_2$ strongly depend on the detuning $\Delta$.

\begin{figure}[htbp]
	\centering
	\resizebox{1\columnwidth}{!}{
		\includegraphics{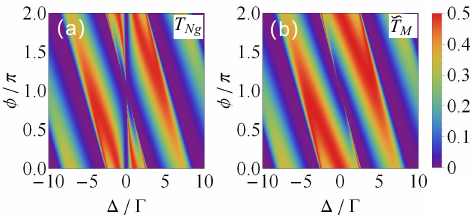}
	}
	\renewcommand\figurename{\textbf{FIG.}}
	\caption[8]{In the non-Markovian regime, the spectra of the single-photon transfer rate $T_{Ng}$ in panel (a) and $\tilde{T}_{M}$ in panel (b) are presented as functions of detuning $\Delta$ and phase $\phi_1^{'} = \phi$. These spectra are calculated with the parameters $\Gamma_1 = \Gamma_3 = \Gamma$, $\Gamma_2 = 0.25\Gamma$, $\Gamma_4 = 0$, and $\tau = 1$.
	}
	\label{fig.8}
\end{figure}

\begin{figure*}[htbp]
	\centering
	\resizebox{1.8\columnwidth}{!}{
		\includegraphics{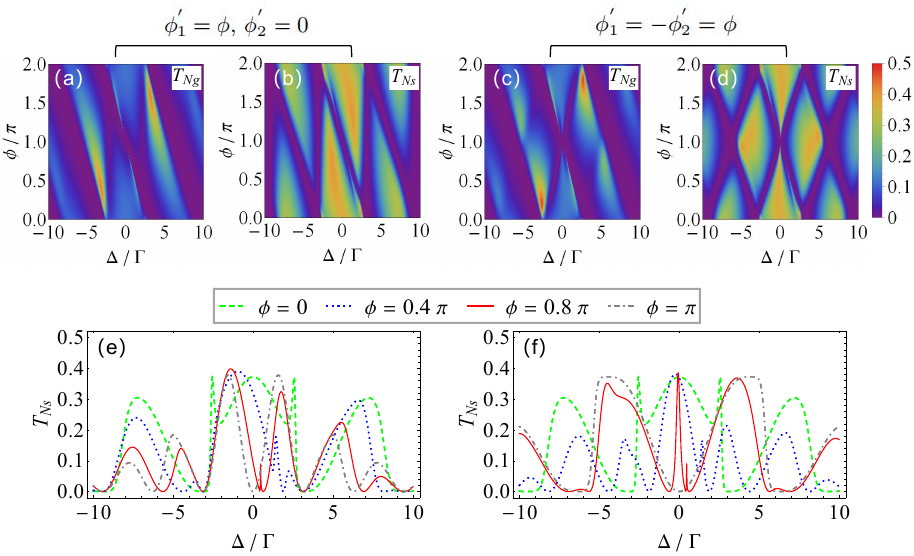}
	}
	\renewcommand\figurename{\textbf{FIG.}}
	\caption[9]{In the non-Markovian regime, the spectra of the single-photon transfer rate $T_{Ng}$ in panels (a) and (c) and $T_{Ns}$ in panels (b) and (d) are shown as functions of detuning $\Delta$ and phase $\phi$. These spectra are calculated with $\phi_1^{'} = \phi$ and $\phi_2^{'}  = 0$ in panels (a) and (b), and $\phi_1^{'}  = -\phi_2^{'}  = \phi$ in panels (c) and (d). Panels (e) and (f) depict the profiles of $T_{Ns}$ for different phase values, corresponding to the cases in panels (b) and (d), respectively. Other parameters are set as $\Gamma_1 = 0.32\Gamma$, $\Gamma_2 = \Gamma_3 = \Gamma_4 = \Gamma$, and $\tau = 1$.}
	\label{fig.9}
\end{figure*}

We first consider single-photon isolation in the non-Markovian regime in Fig. \ref{fig.8}. Compared to the Markovian regime, the most notable difference is that the spectra still exhibit large values when $\phi=\pi$. Specifically, with $\phi_1 = \pi + \tau \Delta$, the numerator in the expressions for $T_{Ng}$ and $\tilde{T}_{M}$ in Eq. (\ref{eq12}) is not always zero due to the term $\tau \Delta$. For instance, when $\phi = \pi$ and $\Delta = 4\Gamma$, we find $T_{Ng} = 0.47$ and $\tilde{T}_{M} = 0.49$. As shown in Fig. \ref{fig.8} (a), the spectrum of $T_{Ng}$ is cut off by the region of $\Delta = 0$.
In Fig. \ref{fig.8} (b), the large value of the transfer rate $\tilde{T}_{M}$ shifts from being concentrated in an $\mathcal{S}$-shaped area to an obtuse triangle-shaped area. However, for $\Delta = 0$, when the phase $\phi \in [0, 0.1\pi]$ or $\phi \in [1.9\pi, 2\pi]$, $\tilde{T}_{M} \approx 0.5$, which is similar to the case in the Markovian regime and suitable for single-photon isolation.

Figure \ref{fig.9} simulates nonreciprocal frequency conversion in the non-Markovian regime. We examine the cases of $\phi_1^{'} = \phi$, $\phi_2^{'} = 0$ in Fig. \ref{fig.9} (a) and (b), and $\phi_1^{'} = -\phi_2^{'} = \phi$ in Fig. \ref{fig.9} (c) and (d). The spectra of the transfer rates are represented as multiple polygons. When adjusting the parameter from $\phi_2^{'} = 0$ to $\phi_2^{'} = -\phi$, the polygons change from inclined to uninclined. Similar to the Markovian regime, an ultranarrow window is generated for both $T_{Ng}$ and $T_{Ns}$ if $\phi \rightarrow \pi$ around $\Delta = 0$.
The detailed description of $T_{Ns}$ is shown in Fig. \ref{fig.9} (e) and Fig. \ref{fig.9} (f). As discussed in Section \ref{IIIA}, at resonance, the proportion of frequency conversion and the transfer rates remain invariant. The case with $\phi = 0.8\pi$ and $\phi_2^{'} = -\phi$ continues to be suitable for frequency conversion. Additionally, compared to the situation shown in Fig. \ref{fig.7} (e) and (f), more peaks are generated at $\Delta \neq 0$, and $T_{Ns}$ becomes more strongly $\Delta$-dependent when $\phi = \pi$ across the entire bandwidth. In the non-Markovian regime, there are more opportunities to explore frequency conversion, including nonreciprocal frequency conversion at off-resonant conditions.

\section{ EXPERIMENTAL FEASIBILITY}\label{IV}

In this section, we provide a detailed analysis of the experimental feasibility of this scheme. The model can be implemented using artificial atoms, which have been instrumental in establishing multiple coupling points to the flux line \cite{kannan2020waveguide}. The coupling strength $g_i$ (analogous to the radiative decay rate $\Gamma_i$) chosen in the main text can be realized in practice. Several methods can be employed to adjust the qubit-waveguide coupling strength: (1) Transmon qubits can be coupled to a superconducting transmission line with interaction mediated by a Josephson loop, and time-dependent couplings can be achieved by modulating the external flux through the loops \cite{PhysRevB.87.134504, PhysRevLett.110.107001, PhysRevA.92.012320, kounalakis2018tuneable}; (2) Capacitor insertion involves placing a capacitor between the voltage nodes of the two circuits involved, allowing the coupling strength to be adjusted by varying the coupling capacitances \cite{PhysRevA.83.063827, krantz2019quantum}; (3) Ultracold atoms in optical lattices can utilize time-dependent couplings by dynamically modulating the relative position of the potentials \cite{PhysRevLett.122.203603}.

The giant $\Lambda$-type atom can be realized in superconducting circuits \cite{PhysRevLett.111.153601, PhysRevLett.113.063604, inomata2016single}. By coupling twice with a transmission line, a giant atom is constructed, with the phases $\phi_1$ and $\phi_2$, induced by the distances between coupling points, being adjustable through modifications in voltages and currents or through electric and magnetic fields \cite{gu2017microwave}. Additionally, such a model can be implemented by coupling a GaAs quantum dot to an optical fiber \cite{davancco2009fiber}. The frequencies $\omega_1/2\pi$ and $\omega_s/2\pi$ can be tuned depending on the strength of the external magnetic field, with values in the ranges of $10^{14}$ Hz and $10^{9}$ Hz, respectively. Specifically, reference values are $\omega_1/2\pi = 3.7 \times 10^{14}$ Hz and $\omega_s/2\pi = 6 \times 10^{14}$ Hz, as proposed in Refs. \cite{PhysRevB.15.816, PhysRevLett.108.103902}.

\section{CONCLUSION}\label{V}

To summarize, we investigate single-photon isolation and nonreciprocal frequency conversion using a two-level atom and a $\Lambda$-type atom, each simultaneously coupled to two separate waveguides. The transmission of single photons is influenced by the coupling strength $g_i$. When $g_4 = 0$, the system acts as a diode, allowing photons to transfer unidirectionally from waveguide $N$ to waveguide $M$. In cases where $g_4 \neq 0$, frequency conversion can occur through the atomic transition $\vert s \rangle \leftrightarrow \vert e \rangle$ for photons incident from waveguide $M$ to waveguide $N$. During this inelastic scattering process, there is an atomic state transition from the initial state $\vert g \rangle$ to the final state $\vert s \rangle$ of the three-level atom. This ultimately results in a change of the photon wave vectors (i.e., $k \rightarrow q$) to satisfy energy conservation. Simultaneously, photons can transfer from waveguide $N$ to waveguide $M$ without changing frequency. This entire process corresponds to nonreciprocal frequency conversion. Additionally, the efficiency of frequency conversion is governed by the parameter $\Gamma_i$. For instance, at resonance, with $\Gamma_2 = \Gamma_3 = \Gamma_4 = \Gamma$ and $\Gamma_1 = 0.32 \Gamma$, the transfer rate $T_{Ns} = 0.37$ and $\eta = 0.76$. In contrast, for $\Gamma_2 = \Gamma_3 = \Gamma$ and $\Gamma_1 = \Gamma_4 = 0.25 \Gamma$, the values are $T_{Ns} = 0.3$ and $\eta = 0.94$. Therefore, it is essential to adjust the parameters according to specific requirements to optimize the system's performance.

We also explore the Markovian and non-Markovian regimes in the giant atom system. Our numerical simulations demonstrate that the distance $d$ between the atomic coupling points, which generates phases $\phi_1$ and $\phi_2$, significantly affects the scattering process of single photons. Specifically, for the case where $\phi_1^{'} = -\phi_2^{'} = \phi$, an effective ultranarrow transmission window is observed, even when $\phi = 0.8\pi$. In the non-Markovian regime, transfer rates exhibit strong dependence on $\Delta$, with spectra concentrating in polygonal regions, and the probability remains non-zero across the entire band when $\phi = \pi$. The non-Markovian regime offers additional opportunities for exploring single-photon isolation and frequency conversion at non-resonant conditions. This work demonstrates the feasibility of achieving optical non-reciprocity at the single-photon level in atom-waveguide systems, with potential applications in quantum communication and quantum information processing.

\begin{figure}[htbp]
	\centering
	\resizebox{0.8\columnwidth}{!}{
		\includegraphics{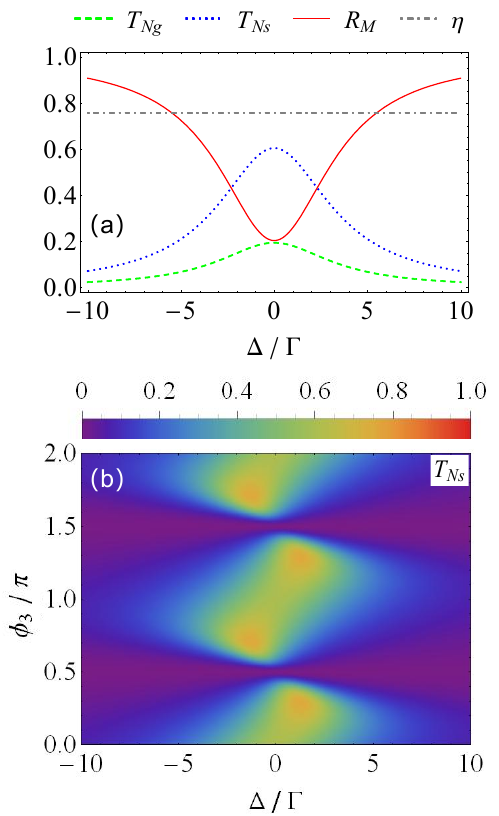}
	}
	\renewcommand\figurename{\textbf{FIG.}}
	\caption[10]{The spectra of the single-photon transfer rates $T_{Ng}$ (green dashed line), $T_{Ns}$ (blue dotted line), reflection rate $R_{M}$ (red solid line), and the proportionality factor $\eta$ (gray dot-dashed line) are plotted as functions of detuning $\Delta$ in panel (a) with $\phi_3 = 0$. In panel (b), the spectra of the single-photon transfer rate $T_{Ns}$ are shown as functions of detuning $\Delta$ and phase $\phi_3$. These spectra are calculated with the parameters $\Gamma_2 = \Gamma_3 = \Gamma_4 = \Gamma$ and $\Gamma_1 = 0.32 \Gamma$.}
	\label{fig.10}
\end{figure}

\section*{ACKNOWLEDGMENTS}

This work is supported by the National Natural Science
Foundation of China under Grant No. 12375018.

\appendix

\section*{APPENDIX: Single photon scattering with one terminated waveguide}\label{Appendix}

In this appendix, we replace the infinite linear lower bus waveguide (i.e., waveguide $M$) in Fig. \ref{fig.1} with a semi-infinite waveguide to investigate the single-photon scattering process. The semi-infinite waveguide can be implemented using a photonic crystal waveguide \cite{lodahl2004controlling} or a microwave transmission line \cite{gu2017microwave,PhysRevLett.112.170501,PhysRevA.101.053861}. To simplify the calculation, we assume that the two atoms couple to the waveguides at the same points.
Considering a single photon incident from waveguide $M$, the wave functions are given respectively by

\begin{equation}\label{eq14}
	\begin{aligned}
		&c_{R,M}(x)=\lbrace\Theta(-x)+A^{'}[\Theta(x)-\Theta(x-L)]\rbrace e^{ikx},\\
		&c_{L,M}(x)=\lbrace r_{1}\Theta(-x)+B^{'}[\Theta(x)-\Theta(x-L)]\rbrace e^{-ikx},\\
		&c_{R,Ng}(x)=t_{4,g}\Theta(x)e^{ikx},\\
		&c_{L,Ng}(x)=t_{3,g}\Theta(-x)e^{-ikx},\\
		&c_{R,Ns}(x)=t_{4,s}\Theta(x)e^{iqx},\\
		&c_{L,Ns}(x)=t_{3,s}\Theta(-x)e^{-iqx}.
	\end{aligned}
\end{equation}
Here, $L$ represents the distance between the coupling point and the end of waveguide $M$. $A^{'}$ and $B^{'}$ are the probability amplitudes for finding right-moving and left-moving photons with wave vector $k$ in the region $0 < x < L$. Based on the eigenvalue equation and the hard-wall boundary condition $c_{R,M}(L)+c_{L,M}(L)=0$
, the following expressions are obtained:

\begin{equation}\label{eq15}
 \begin{aligned}
    &t_{3,g}=t_{4,g}=\dfrac{-(1+e^{2i\phi_3})\chi_4    \sqrt{\Gamma_1\Gamma_3}}{\chi_4[\chi_3+(1+e^{2i\phi_3})\Gamma_1]+(1+e^{2i\phi_3})\chi_3\Gamma_2},\\
    &t_{3,s}=t_{4,s}=\dfrac{-(1+e^{2i\phi_3})\chi_3    \sqrt{\Gamma_2\Gamma_4}}{\chi_4[\chi_3+(1+e^{2i\phi_3})\Gamma_1]+(1+e^{2i\phi_3})\chi_3\Gamma_2},\\
 \end{aligned}
\end{equation}
with
\begin{equation}\label{eq16}
\begin{aligned}
&\chi_3=\Gamma_3-i\Delta,\\	
&\chi_4=\Gamma_4-i\Delta,\\
\end{aligned}
\end{equation}
where $\phi_3=kL$.

Figure \ref{fig.10} (a) depicts the transfer rates $T_{Ng}$, $T_{Ns}$, reflection rate $R_{M}$, and proportionality factor $\eta$ as functions of detuning $\Delta$ with $\phi_3=0$. In this configuration, no photons are transmitted to the other side of waveguide $M$ and are reflected by the termination. Consequently, the transfer rates are improved compared to the case in Section \ref{IIB}. Specifically, at $\Delta=0$, $T_{Ng}=0.19$ and $T_{Ns}=0.6$, with $T_{Ns}$ exceeding the limitation of a transfer rate less than 0.5. We also examine the influence of phase $\phi_3$; as shown in Fig. \ref{fig.10}, the transfer rate $T_{Ns}$ varies periodically with $\phi_3$. Frequency conversion is not achievable at $\phi_3=0.5\pi$ and $1.5\pi$ across the entire bandwidth. Additionally, the large transfer rates are concentrated in the range $\Delta \in [-3\Gamma, 3\Gamma]$.

\end{document}